\documentclass[conference]{IEEEtran}
\IEEEoverridecommandlockouts
\usepackage{cite}
\usepackage{amsmath,amssymb,amsfonts}
\usepackage{algorithmic}
\usepackage{graphicx}
\usepackage{textcomp}
\usepackage{xcolor}
\def\BibTeX{{\rm B\kern-.05em{\sc i\kern-.025em b}\kern-.08em
    T\kern-.1667em\lower.7ex\hbox{E}\kern-.125emX}}
\begin{document}

\title{Blockchain-Enabled IoV: Secure Communication and Trustworthy Decision-Making\\
\thanks{J. Sun, Q. Shi and G. Jin are with College of Electronic and Information Engineering, Tongji University, Shanghai 201804, PR China, E-mail: {2252086, qishi, 2252540}@tongji.edu.cn; H.Xu and E. Liu are with College of Electronic and Information Engineering and Shanghai Engineering Research Center for Blockchain Applications And Services, Tongji University, Shanghai, China, E-mail: {hao.xu, erwu.liu}@ieee.org.}
}

\author{Jingyi Sun, Qi Shi, Guodong Jin, Hao Xu and Erwu Liu
}

\maketitle

\begin{abstract}
The Internet of Vehicles (IoV), which enables interactions between vehicles, infrastructure, and the environment, faces challenges in maintaining communication security and reliable automated decisions. This paper introduces a decentralized framework comprising a primary layer for managing inter-vehicle communication and a sub-layer for securing intra-vehicle interactions. By implementing blockchain-based protocols like Blockchain-integrated Secure Authentication (BiSA) and Decentralized Blockchain Name Resolution (DBNR), the framework ensures secure, decentralized identity management and reliable data exchanges, thereby supporting safe and efficient autonomous vehicle operations.
\end{abstract}


\section{Introduction}
The Internet of Vehicles (IoV) represents the convergence of vehicles with digital technologies, enabling communication and interaction between vehicles, infrastructure, pedestrians, and the broader environment. IoV is an essential component of smart transportation systems, facilitating Vehicle-to-Everything (V2X) communications, including Vehicle-to-Vehicle (V2V), Vehicle-to-Infrastructure (V2I), and Vehicle-to-Pedestrian (V2P) interactions. IoV systems must meet several stringent requirements to function effectively and safely. These include low latency, high availability, robust security, and efficient communication protocols. As IoV systems become more prevalent, the need for secure communication and trustworthy decision-making processes becomes increasingly critical.

\section{Challenges in IoV: Communication Security and Trustworthy Decision-Making}
Despite its potential, IoV faces significant challenges that hinder its full deployment. Two of the most pressing issues are communication security and the trustworthiness of decision-making processes.

\subsection{Communication Security}

The Internet of Vehicles (IoV) and Vehicular Ad-Hoc Networks (VANETs) are vulnerable to various security attacks and threats, posing significant challenges to the communication between vehicles and infrastructure. Common attacks include Sybil attacks, Denial of Service (DoS) attacks, Distributed Denial of Service (DDoS) attacks, Black Hole attacks, Grey Hole attacks, Node Impersonation attacks, etc \cite {b1}.

These attacks not only threaten the security of communication between vehicles and infrastructure but also potentially lead to traffic accidents, privacy breaches, and loss of vehicle control.

\subsection{Trustworthy Decision-Making}

The reliability of decision-making in IoV is another critical concern. As IoV systems often rely on automated decision-making, ensuring that these decisions are based on accurate and secure data is essential. Unreliable decision-making can result in incorrect responses to traffic conditions, leading to accidents, traffic congestion, and other adverse outcomes. A single malicious message possesses the capability to compromise the entire network, thereby putting human lives at risk on the road \cite {b2}. 

The lack of a secure framework for managing and validating the data used in these decisions exacerbates the problem, making IoV systems susceptible to manipulation and errors.

\section{Blockchain-Enabled Solutions for IoV}

To address these challenges, blockchain technology offers a promising solution by providing a decentralized framework for secure communication and trustworthy decision-making in IoV systems. A blockchain-enabled approaches have been proposed to enhance the security and reliability of IoV systems.

\begin{figure}
    \centering
    \includegraphics[width=1\linewidth]{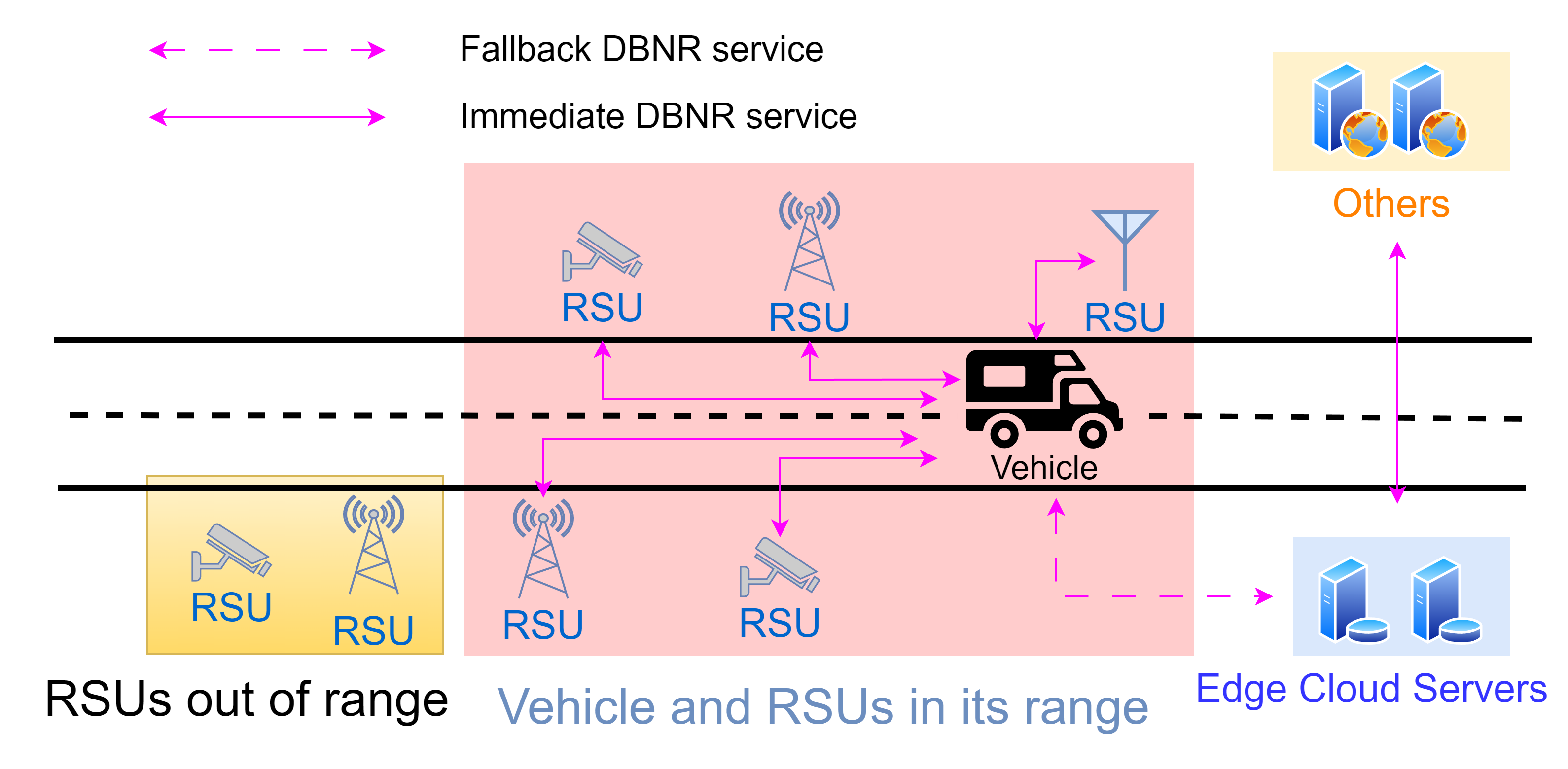}
    \caption{Overall structure of the outer layer.}
    \label{fig:enter-label}
\end{figure}
\begin{figure}
    \centering
    \includegraphics[width=1\linewidth]{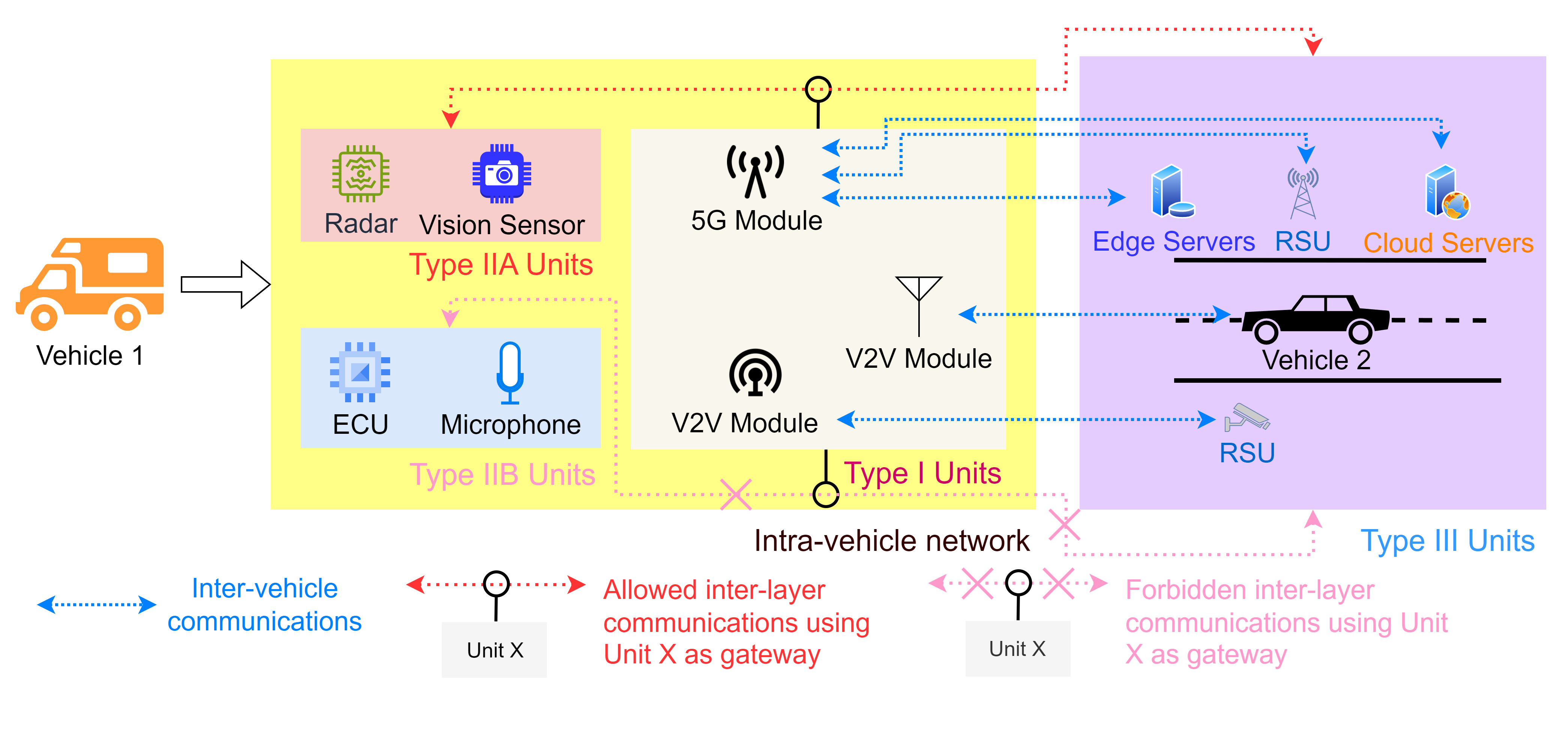}
    \caption{Overall structure of the inner layer.}
    \label{fig:enter-label}
\end{figure}
\subsection{Structure of Blockchain-enabled Decentralized Framework}
The blockchain-enabled decentralized framework is designed to enhance the security and reliability of IoV systems by decentralizing identity management and securing communication channels across the network.

The framework is structured into two layers, which represent different communication contexts: the Primary Layer manages inter-vehicle communication identities and routing information, and the Sub-layer handles intra-vehicle communications, ensuring secure communication between devices within a vehicle. \cite {b3} \cite {b4}. While these layers operate autonomously to handle specific tasks, they also work in tandem to ensure that communication security is guaranteed while vehicles exchange data that facilitates autonomous driving and other sophisticated functions. 

\subsubsection{The Primary Layer}

As shown in Fig. 1, this layer is comprised of Road Side Units (RSUs), edge servers, cloud servers, and vehicles. The primary function of this layer is to manage and secure inter-vehicle communication. Each participant is assigned a unique identity managed and verified by a blockchain-based system embedded in RSUs and edge servers. By introducing BiSA (Blockchain-integrated Secure Authentication), vehicles send requests regarding identity authentication and the topological addresses of devices involved in communication to RSUs and edge servers. These requests are processed redundantly using a modified Practical Byzantine Fault Tolerance (PBFT) algorithm. Once communication identities are successfully authenticated, encrypted connections are established, resilient against threats such as Man-in-the-Middle (MITM) attacks and spoofing. This ensures secure V2V, V2I, and V2P communications. 

\subsubsection{The Sub-layer}

As shown in Fig. 2, each vehicle has its own sub-layer that manages intra-vehicle communication, ensuring that interactions between internal components (such as sensors and control units) remain secure. Through DBNR (Decentralized Blockchain Name Resolution), the sub-layer bridges with the primary layer, allowing non-sensitive components within the vehicle to safely communicate with other vehicles and infrastructure while isolating security- and privacy-sensitive components. This guarantees secure intra-vehicle communication while also supporting safe data exchanges needed for autonomous driving and other advanced functions.

\subsection{Solutions of the Framework to the Aforementioned Challenges}

\subsubsection{Ensuring Communication Security} The system introduces a Blockchain-integrated Secure Authentication (BiSA) as a core protocol for securing communications both within and between vehicles \cite{b5}. BiSA achieves communication encryption by using blockchain-based identity management and mutual authentication, where two parties verify each other's identities through their blockchain addresses without relying on third-party authorities. After successful authentication, a session key is generated and used for end-to-end encrypted communication. By employing BiSA, the system ensures that all communications are authenticated and encrypted, providing robust protection against unauthorized access, eavesdropping, and other cyber threats.

\subsubsection{Ensuring Trustworthy Decision-Making} Decentralized Blockchain Name Resolution (DBNR) plays a crucial role in the issue of trustworthy decision-making by providing a decentralized, immutable, and reliable system for managing the identities of all communication entities \cite{b6}.

DBNR integrates into RSUs and edge servers, where it manages the mapping of blockchain addresses to their corresponding network interface identifiers and topological locations. This ensures that all routing information is secure, up-to-date, and immutable, thereby preventing malicious alterations or spoofing of identity data.

\section{Conclusion}
As IoV systems continue to evolve, ensuring secure communication and trustworthy decision-making is essential for their successful deployment. Blockchain technology provides a robust framework for addressing these challenges by decentralizing identity management and securing communication channels, thereby mitigating the risks associated with traditional, centralized systems. By integrating blockchain into IoV infrastructure and decision-making processes, these systems can achieve greater security, reliability, and efficiency, paving the way for the next generation of smart transportation solutions.

\vspace{12pt}
\color{red}
\end{document}